# Crossover of conduction mechanism in $Sr_2IrO_4$ epitaxial thin films

Chengliang Lu[1,2], Andy Quindeau[1], Hakan Deniz[1], Daniele Preziosi[1], Dietrich Hesse[1], and Marin Alexe[1,3*]

[1] *Max Planck Institute of Microstructure Physics, Weinberg 2, D-06120 Halle (Saale), Germany*

[2] *School of Physics, Huazhong University of Science and Technology, Wuhan 430074, China*

[3] *Department of Physics, Warwick University, Coventry CV4 7AL, United Kingdom.*

Abstract

High quality epitaxial $Sr_2IrO_4$ thin films with various thicknesses (9-300 nm) have been grown on $SrTiO_3$ (001) substrates, and their electric transport properties have been investigated. All samples showed the expected insulating behavior with a strong resistivity dependence on film thickness, that can be as large as three orders of magnitude at low temperature. A close examination of the transport data revealed interesting crossover behaviors for the conduction mechanism upon variation of thickness and temperature. While Mott variable range hopping (VRH) dominated the transport for films thinner than 85 nm, high temperature (>200 K) thermal activation behavior was observed for films with large thickness (≥85 nm), which was followed by a crossover from Mott to Efros-Shklovskii (ES) VRH in the low temperature range. This low temperature crossover from Mott to ES VRH indicates the presence of a Coulomb gap (~3 meV). Our results demonstrate the competing and tunable conduction in $Sr_2IrO_4$ thin films, which in turn would be helpful for understanding the insulating nature related to strong spin-orbital-coupling of the 5*d* iridates.

---

[*] Corresponding author: M.Alexe@warwick.ac.uk.

In recent years, oxides with the 5$d$-element Ir have been established as a fertile ground for studying new physics arising from the large relativistic spin-orbit-coupling (SOC) [1-3]. So far, a variety of exotic properties related to the strong SOC such as intriguing insulating behavior [3-10], giant magnetoelectricity [11], colossal magnetoresistance driven by spin-lattice coupling [12], high temperature superconductivity [13], correlated topological insulator [14], etc. have been experimentally observed or theoretically predicted in the iridates. Among them, the insulating behavior observed in many iridates stands out since a metallic state would be expected if simply considering the interplay between bandwidth and Hubbard repulsion as in 3$d$ transition metal oxides [15].

The single-layer iridate $Sr_2IrO_4$ (SIO), which is isostructural to the undoped high-$T_C$ cuprate $La_2CuO_4$ and the $p$-wave superconductor $Sr_2RuO_4$, has been subjected to the most extensive investigations on the insulating nature. A picture of a SOC induced Mott insulator was developed recently based on theoretical calculations and spectroscopic studies, in which the band-gap opening was ascribed to strong electron-electron coupling [1]. Although this picture has been widely accepted since 2008, very recent experimental and theoretical studies proposed a Slater mechanism for the gap opening in SIO, bringing this topic under debate [16-18].

In addition to the newly raised controversy on the insulating nature of SIO, the conduction mechanism of the insulating state has been a long-standing issue. In an early work, variable range hopping (VRH) conduction was reported in single crystalline SIO [3]. A recent study also focusing on SIO single crystals revealed that the conduction well follows the thermal activation mechanism with an energy gap of $\Delta \sim 107$ meV perfectly matching the value of optical measurements [1, 12]. Further investigations of the same group showed a VRH dominated conduction in doped SIO even though the doping content was small [19, 20]. In $Ba_2IrO_4$ (BIO), although it highly resembles SIO in terms of structural and electronic properties, good VRH conduction was observed in the whole measured temperature ($T$) range. Moreover this type of conduction is robust against carrier doping [6], which is very different from the SIO case. In fact, a close examination of the conduction mechanism in SIO and BIO can provide some more interesting details. While the doped SIO shows 3-dimentional (3D) VRH conduction [19, 20], the transport of BIO and its derivates well follows a 2D mechanism [6]. Significant Anderson localization with noticeable Coulomb interaction was identified in $Sr_2(Ir_{1-x}Rh_x)O_4$ [20], but simple VRH conduction without any sign of electron-electron interaction was detected in electron-doped $Sr_2IrO_{4-\delta}$ [19], which could probably be related to the SOC. These show intriguing conduction behavior in the single layered iridates, pointing to a rather poor understanding of the conduction mechanism in SIO. Moreover, this issue became even more puzzling in the case of SIO thin films, in which the thermal activation model was frequently used to determine the energy gap of SIO, although a linear fit on a plot of ln$\rho$-1/$T$ is questionable [21, 22]. Therefore, it is of high interest and actuality to comprehensively investigate the conduction mechanism of these materials, which might in turn be helpful for understanding the unexpected insulating state of the 5$d$ iridates.

In the present work, we study the thickness ($t$) dependence of the conduction mechanism in epitaxial SIO thin films grown on $SrTiO_3$ (STO) (001) substrates. The reason for choosing the STO (lattice parameter $a$=3.905 Å) as a substrate is the good lattice fit, i.e. only 0.4% misfit, with SIO ($a$=3.888 Å) [23]. In addition, both STO and SIO have equivalent basal plane lattice parameters ($a$=$b$), preventing undesirable anisotropic strain. We vary the film thickness from 9 nm to 300 nm in order to explore possible correlations between ultrathin film- and bulk-like conduction. Our experimental results reveal interesting crossovers of the conduction mechanism as temperature and thickness are changed. While Mott-VRH [24] dominates the conductivity in very thin films, a crossover from bulk-like high-$T$ thermal activation to low-$T$ Efros-Shklovskii (ES) VRH [25] is evidenced in films with large thickness.

Epitaxial SIO films with various thicknesses ranging from 9 nm to 300 nm were grown on STO (001) substrates with vicinal surface using pulsed laser deposition (PLD). A KrF excimer laser (248 nm wavelength) was used for the deposition with a repetition rate of 1 Hz, and the laser energy density was set at ~0.6 J/cm$^2$. A stoichiometric SIO target was used for the ablation. The depositions were performed at a substrate temperature of 830 ºC and 0.05 mbar of oxygen partial pressure. After deposition, the samples were cooled down in 1 atm of oxygen pressure to avoid the formation of oxygen vacancies. X-ray diffraction (XRD) analyses were carried out using a Philips X'Pert diffractometer. The surface morphology of the samples was investigated using a Digital Instrument D-5000 atomic force microscope (AFM). Transmission electron microscopy (TEM) and electron diffraction studies were performed on a Philips CM20T microscope at 200 kV, and atomic resolution high angle annular dark-field scanning transmission electron microscopy (HAADF-STEM) was carried out using a FEI TITAN 80-300 microscope.

Transport measurements were performed using a standard four-probe method in a physical property measurement system (PPMS, Quantum Design).

Fig. 1 (a) shows $\theta$-$2\theta$ scans of samples with $t$=9 nm, 18 nm, and 300 nm. All the samples show a pure phase and a single c-axis orientation. Although the film thickness is increased largely, there is no visible shift in relevant diffraction peaks, indicating insignificant strain relaxation due to the increase in film thickness. This is consistent with the small lattice misfit (0.4%) between SIO and STO. The close $c$ value of both thin film (12.74 Å) and bulk SIO (12.90 Å) [23] confirms the modest in-plane tensile strain of the films. Fig. 1(c) presents the atomic force microscopy image of the film with $t$=18 nm. A clear terrace surface structure can be seen, indicating a good quality of the film. In order to capture more details of the crystalline structure, TEM cross-section observations along the [010] zone axis of the film with $t$=18 nm were conducted, as shown in the right of Fig. 1(b). The corresponding electron diffraction pattern is presented on the left of Fig. 1(b), and all the diffraction spots can be well indexed to the film or the substrate lattice, respectively. This reveals a good epitaxy with respect to the substrate. The high resolution STEM image shown in Fig.1 (d) further confirms the good epitaxy and quality of the sample with $t$=18 nm. SrO and $IrO_2$ planes stack in ABA-ABA sequence across the film, and the $c$ value is estimated to be ~12.76 Å matching well the value obtained from XRD analysis.

Figure 2 displays in-plane resistivity ($\rho$) as a function of $T$ for all the samples with different thicknesses ranging from 9 nm to 300 nm. As expected, all the samples show insulating behavior in the whole measured $T$ range, which is consistent with earlier reports on SIO/STO thin films [21, 26]. Interestingly, by significantly varying the film thickness, the dependence of $\rho$ with $T$ is changed drastically. In the same time, $\rho$ decreases monotonically as $t$ increases, at low temperature, the difference in $\rho$ can be as large as three orders of magnitude. This feature strongly suggests the significance of size effects on the electric transport of SIO/STO films, and probably indicates different conduction mechanism, which will be discussed below.

The resistivity data are firstly analyzed using the thermal activation model $\rho \sim \exp(\Delta/2k_{B}T)$, where $\Delta$ is the thermal activation energy and $k_{B}$ is Boltzmann's constant. This model can well describe the transport of bulk SIO as aforementioned [12]. In Fig. 3(a), however, the fitting quality is really poor for films thinner than 85 nm, indicating that the transport can not be described by thermal activation across a single band gap. Actually, a similar very limited linear fitting region in the Arrhenius plot (ln$\rho$ vs 1/$T$) was also revealed in earlier works on SIO thin films with thickness less than 60 nm [21, 22]. However, with increasing the film thickness to $t$ ≥85 nm, a good linear relationship can be seen in the Arrhenius plot, especially in the high-$T$ range (>200 K), suggesting a thermal activation conduction in the films which is similar to the case in SIO bulk crystals. The activation energy $\Delta$ ranges from 83.8 to 97.5 meV, which is also close to the value of bulk SIO crystals (~107 meV) [12].

Instead of the thermally activated conduction observed in films with large thickness, the transport data of films with small thickness fit well to the relation characteristic of the 3D Mott-VRH mechanism [24, 27]:

$$\rho = \rho_0 \exp\left(\frac{T_M}{T}\right)^{\frac{1}{4}} \quad (1)$$

where $\rho_0$ is the resistivity coefficient, and $T_M$ is the characteristic temperature. As shown in Fig. 3(b), films with $t$<85 nm show two distinct, high and low, temperature ranges where ln$\rho$ shows very good linear behavior with different slopes, suggesting two distinct values of $T_M$ ($T_{M1}$ for high-$T$, and $T_{M2}$ for low-$T$). However, this Mott-VRH dominated transport is suppressed for both high-$T$ and low-$T$ ends for the films with $t$≥85 nm, which is marked by a clear deviation from the linear fitting. The deviation in the high-$T$ region can be reasonably explained by the emergence of thermal activation behavior, while the nonlinear relationship of ln$\rho$-$T^{-1/4}$ in the low-$T$ region is intriguing and will be discussed later.

According to the Mott theory on VRH, the average hopping distance $R_M$ must be larger than the localization length $a$, which is [24, 27]:

$$\frac{R_M}{a} = \frac{3}{8}\left(\frac{T_M}{T}\right)^{\frac{1}{4}} > 1 \quad (2)$$

The ratio $R_M/a$ derived from the fitting in Fig. 3(b) is larger than 2 for all the samples within the temperature range where we consider the Mott-VRH to be dominant, which satisfies the criterion for Mott-VRH. The obtained values of $R_M/a$ are listed in Table I. From the fitting, it is also possible to estimate the density of states $N(E_F)$ at the Fermi level and the localization length $a$ according to: [24, 27]

$$T_M = \frac{18}{k_B a^3 N(E_F)} \quad (3)$$

In (3), it is required to know at least one parameter (either $a$ or $N(E_F)$) for estimating the other one. Based on the reported heat capacity data of bulk SIO [20], we can assume $N(E_F)$ to be $\sim 10^{47}/Jm^3$. Therefore, the calculated localization length $a$ of our samples ranges from 0.5 Å to 3.9 Å, which is comparable with the Ir-O bond length ($\sim 2$ Å) of SIO, verifying once more the validity of the Mott-VRH mechanism. In addition, the fitting also gives the Mott hopping energy: [24, 27]

$$E_M = \frac{1}{4} k_B T \left( \frac{T_M}{T} \right)^{\frac{1}{4}} \quad (4)$$

which is in the order of few tens of meV. According to equation (2), the average hopping distance $R_M$ is estimated to be at least an order of magnitude smaller than the film thickness, consisting with the 3D hopping conduction [27].

For the low-$T$ transport data in the films with large thickness such as $t \geq 85$ nm neither the thermal activation model nor the Mott-VRH mechanism are valid. For this specific film thicknesses and temperature range we use the ES-VRH model: [25, 27]

$$\rho = \rho_0 \exp\left( \frac{T_{ES}}{T} \right)^{\frac{1}{2}} \quad (5)$$

where $T_{ES}$ is the characteristic temperature of the ES-VRH mechanism. Indeed, a very good linear relationship of $\ln\rho$ vs. $T^{-1/2}$ can be seen in the low-$T$ region for the samples with $t \geq 85$ nm (Fig. 3(c)). The linear fitting range becomes larger as $t$ increases, suggesting that the low-$T$ transport of the films with $t \geq 85$ nm are dominated by the ES-VRH mechanism. For the ES-VRH theory to be valid, the average hopping distance $R_{ES}$ must be larger than the localization length $a$, and the ratio $R_{ES}/a$ can be calculated according to:

$$\frac{R_{ES}}{a} = \frac{1}{4}\left( \frac{T_{ES}}{T} \right)^{\frac{1}{2}} \quad (6)$$

and should be larger than unity [25, 27]. All $R_{ES}/a$ values for the films with $t \geq 85$ nm are larger than 1, satisfying thus the criterion of the ES-VRH model. The ES hopping energies calculated according to $E_{ES} = 1/2 k_B T (T_{ES}/T)^{1/2}$ [25, 27] are listed in Table II.

From the above transport data analysis, we note that the Mott-VRH mechanism dominates the transport within the entire temperature range for the samples with $t \leq 45$ nm, while a clear crossover from the Mott to ES VRH conduction mechanism was observed for films with

t≥85 nm. The temperature range at which this crossover occurs is ~20-40 K. This crossover indicates the opening of a Coulomb gap at low temperature. The value of the Coulomb gap can be estimated from: [27, 28]

$$\Delta_{CG} \approx k_B \left(\frac{T_{ES}^3}{T_M}\right)^{1/2} \tag{7}$$

The calculated value of $\Delta_{CG}$ is ~3 meV for all the three samples, which is close to the values obtained in other materials showing a similar crossover feature of the conduction [27, 28]. Regarding the onset of the crossover from Mott to ES VRH, one would expect that the Mott hopping energy is comparable to the ES hopping energy ($E_M = 1/4 k_B T (T_M/T)^{1/4} = E_{ES} = 1/2 k_B T (T_{ES}/T)^{1/2}$), and hence we can calculate the temperature where the crossover arises according to $T_{cross}=16 T_{ES}^2/T_M$. The calculated values of $T_{cross}$ (listed in Tab. II) are very close to the values obtained from the experimental data. Here, we note that the low-$T$ transport of the thinnest sample ($t$=9 nm) in the present work can be well fitted with both Mott and ES VRH models as shown in Fig. 3(b) and (c). However, the fitting with ES-VRH gives a very large Coulomb gap (~23.7 meV) but very small $T_{cross}$ (7 K) which is far smaller than the experimental $T$ value (180 K), suggesting that the low-$T$ transport of the sample is dominated by Mott hopping and not by ES.

To visualize the results shown above, Figure 4 schematically summarizes the temperature and thickness dependence of the conduction mechanism in epitaxial SIO films. Here regions with different conduction mechanisms as well as the crossover between the different conduction mechanisms are schematically drawn. For the films with $t$≤45 nm, the transport is dominated by the Mott-VRH mechanism, suggesting the existence of localized states in the band gap and a possible finite density of states at the Fermi level. Here we would propose three factors as the possible origins of such VRH conduction: i) the native defects like Sr or Ir vacancies in the present case, which commonly exist in thin films; ii) the strain effect which could modify the band structure of SIO, and iii) possible different transport properties at the interface layer and film surface. For the first factor, since all the samples were synthesized under the same parameters, similar density of the vacancies would be expected in the films with different thickness. Therefore, the contribution of Sr or Ir vacancies to the VRH conduction should not be significantly different for all the samples. The strain has an important effect on the transport through the band gap, but an earlier optical study revealed that the band gap of SIO films is rather strain-independent [29]. Moreover, in our case the lattice misfit between the substrate and SIO is very small (0.4%). So, a significant strain effect on the VRH conduction mechanism is not expected. In Table I, we note that the characteristic temperatures $T_{M1}$ and $T_{M2}$ show a two orders of magnitude decrease as $t$ increases from 9 to 45 nm, and then a moderate variation upon further increasing $t$ to 300 nm. Based on this we might speculate on an important contribution arising from the interface layer or film surface to the transport. For the films with $t \geq 85$ nm, thermal activation dominates the high-$T$ transport. The estimated values of the thermal activation energy (83.8≤Δ≤97.5 meV) are in this case quite close to the value of SIO bulk crystals [12], suggesting a dominating role of the bulk-like conduction in the thick films. In addition, the bulk-like conduction appears concurrently with the Coulomb gap opening, which confirms the picture of Kim *et al.* that SIO is a Mott insulator [1]. The $T$-driven crossover of conduction mechanism of film with $t$=150 nm is plotted in Fig. 5(a), in which the variation of conduction mechanism with $T$ can be clear seen, similar to that of $SnO_2$ nanobelt [30]. Regarding the feature of two Mott-VRH regions in films with small thickness, we further check that using a formula ($w$=d[ln$\sigma$]/d$T$=$pT_0^p/T^{p+1}$) suggested by Zabodskii, where $\sigma$ is the conductivity, $p$ is the critical exponent of the hopping mechanism, and $T_0$ corresponds to $T_M$ or $T_{ES}$ depending on the conduction mechanism [31, 32]. As shown in Fig. 5(b), the obtained exponents are quite close to 0.25, confirming the Mott-VRH conduction in the film for both high- and low-$T$ regions.

The electric transport of epitaxial $Sr_2IrO_4$ thin films grown on $SrTiO_3$ (001) substrates has been investigated in a very large thickness range (9-300 nm). Insulating behavior was observed in all films in the entire temperature range (5-360 K). Resistivity and the conduction mechanism show significant dependence on both thickness and temperature. Pure Mott variable range hopping conduction was revealed in the films with small thickness (≤45 nm). However, in the films with large thickness (85-300 nm), the high temperature transport was dominated by thermal activation, and at low temperature a crossover from Mott to ES variable range hopping conduction occurs. The present study show significant size effects in conduction in epitaxial $Sr_2IrO_4$ thin films, which might be significant for the ongoing debate on the insulating state in iridates.

**Acknowledgement:** This work was partly funded by DFG via SFB 762. C.L.L. acknowledges the funding from the Alexander von Humboldt Foundation and the National Natural Science Foundation of China (Grant Nos. 11104090 and 11374112). M. A. acknowledges the Wolfson research merit award of the Royal Society. Valuable discussions with Prof. James F. Scott are also acknowledged.

*References:*

1. B. J. Kim, H. Jin, S. J. Moon, J.-Y. Kim, B.-G. Park, C. S. Leem, J. Yu, T. W. Noh, C. Kim, S.-J. Oh, J.-H. Park, V. Durairaj, G. Cao, and E. Rotenberg, Phys. Rev. Lett. **101**, 076402 (2008).
2. B. J. Kim, H. Ohsumi, T. Komesu, S. Sakai, T. Morita, H. Tagaki, T. Arima, Science **323**, 1329 (2009).
3. G. Cao, J. Bolivar, S. McCall, J. E. Crow, and R. P. Guertin, Phys. Rev. B **57**, 11039(R) 1998.
4. O. B. Korneta, S. Chikara, S. Parkin, L. E. De Long, P. Schlottmann, and G. Cao, Phys. Rev. B **81**, 045101 (2010).
5. S. M. Disseler, C. Dhital, A. Amato, S. R. Giblin, C. de la Cruz, S. D. Wilson, and M. J. Graf, Phys. Rev. B **86**, 014428 (2012).
6. H. Okabe, M. Isobe, E. T. Muromachi, N. Takeshita, and J. Akimitsu, Phys. Rev. B **88**, 075137 (2013).
7. C. Dhital, S. Khadka, Z. Yamani, C. de la Cruz, T. C. Hogan, S. M. Disseler, M. Pokharel, K. C. Lukas, W. Tian, C. P. Opeil, Z. Wang, and S. D. Wilson, Phys. Rev. B **86**, 100401(R) (2012).
8. G. Cao, V. Durairaj, S. Chikara, S. Parkin, P. Schlottmann, Phys. Rev. B **75**, 134402 (2007).
9. S. Singh and P. Gegenwart, Phys. Rev. B **82**, 064412 (2010).
10. H. C. Lei, W. G. Yin, Z. C. Zhong, and H. Hosono, Phys. Rev. B **89**, 020409(R) (2014).
11. S. Chikara, O. Korneta, W. P. Crummett, L. E. Delong, P. Schlottmann, and G. Cao, Phys. Rev. B **80**, 140407R (2009).
12. M. Ge, T. F. Qi, O. B. Korneta, D. E. De Long, P. Schlottmann, W. P. Crummett, and G. Cao, Phys. Rev. B **84**, 100402(R) (2011).
13. F. Wang and T. Senthil, Phys. Rev. Lett. **106**, 136402 (2011).
14. D. A. Pesin and L. Balents, Nat. Phys. **6**, 376 (2010).
15. M. Iamada, A. Fujimori, and Y. Tokura, Rev. Mod. Phys. **70**, 1039 (1998).
16. R. Arita, J. Kunes, A. V. Kozhevnikov, A. G. Eguiluz, and M. Imada, Phys. Rev. Lett. **108**, 086403 (2012).

17. Q. Li, G. Cao, S. Okamoto, J. Yi, W. Lin, B. C. Sales, J. Yan, R. Arita, J. Kunes, A. V. Kozhenikov, A. G. Eguiluz, M. Imada, Z. Gai, M. Pan, and D. G. Mandrus, Sci. Rep. **3** 3073 (2013).
18. D. Hsieh, F. Mahmood, D. H. Torchinsky, G. Cao, and N. Gedik, Phys. Rev. B **86**, 035128 (2012).
19. O. B. Korneta, T. F. Qi, S. Chikara, S. Parkin, L. E. De Long, P. Schlottmann, and G. Cao, Phys. Rev. B **82**, 115117 (2010).
20. T. F. Qi, O. B. Korneta, L. Li, K. Butrouna, V. S. Cao, X. G. Wan, P. Schlottmann, R. K. Kaul, and G. Cao, Phys. Rev. B **86**, 125105 (2012).
21. C. R. Serrao, J. Liu, J. T. Heron, G. S. Bhalla, A. Yadav, S. J. Suresha, R. J. Paull, D. Yi, J. H. Chu, M. Trassin, A. Vishwanath, E. Arenholz, C. Frontera, J. Zelezny, T. Jungwirth, X. Marti, and R. Ramesh, Phys. Rev. B **87**, 085121 (2013).
22. J. Nichols, O. B. Korneta, J. Terzic, L. E. De Long, G. Cao, J. W. Brill, and S. S. A. Seo, Appl. Phys. Lett. **103**, 131910 (2013).
23. Q. Huang, J. L. Soubeyroux, O. Chmaissem, I. N. Sora, A. Santoro, R. J. Cava, J. J. Krajewski, and W. F. P. Jr. J. Solid State Chem. **112**, 355 (1994).
24. N. F. Mott, J. Non-Cryst. Solids **1**, 1 (1968).
25. A. L. Efros and B. I. Shklovskii, J. Phys. C: Solid State Phys. **8**, L49 (1975).
26. L. Miao, H. Xu, and Z. Q. Mao, Phys. Rev. B **89**, 035109 (2014).
27. R. Rosenbaum, Phys. Rev. B **44**, 3599 (1991).
28. X. Y. Zhang, J. S. Chawla, B. M. Howe, and D. Gall, Phys. Rev. B **83**, 165205 (2011).
29. J. Nichols, J. Terzic, E. G. Bittle, O. B. Korneta, L. E. Delong, J. W. Brill, G. Cao, and S. S. A. Seo, Appl. Phys. Lett. **102**, 141908 (2013).
30. E. R. Viana, J. C. González, G. M. Ribeiro, and A. G. de Oliveira, Phys. Status Solidi RRL **6**, 262 (2012).
31. J. C. González, G. M. Ribeiro, E. R. Viana, P. A. Fernandes, P. M. P. Salomé, K. Gutiérrez, A. Abelenda, F. M. Matinaga, J. P. Leitão, and A. F. da Cunha, J. Phys. D: Appl. Phys. **46**, 155107 (2013).
32. A. G. Zabodskii, Phys.-Usp. **41**, 722 (1998).

*Figure captions:*

Figure 1. (a) X-ray diffraction patterns of the films with thickness $t$=9 nm, 18 nm, and 300 nm. (b) Transmission electron microscopy cross-section image (right) of the 18 nm SIO film (the film is indicated by two dashed lines), and the corresponding electron diffraction pattern (left) where the spots are indexed for the film and substrate, accordingly. The subscripts, 'S' and 'F', denote the substrate and film, respectively. (c) Surface morphology image of the same film in (b). (d) Scanning transmission electron microscopy image of the same film. Inset: enlarged zoom with highlighted Sr (blue dots) and Ir (red dots) atoms.

Figure 2. Temperature dependence of the resistivity of SIO epitaxial films with different thicknesses ranging from 9 nm to 300 nm.

Figure 3. ln$\rho$ versus (a) $1/T$, (b) $T^{-1/4}$, and (c) $T^{-1/2}$ for all investigated films. The solid lines are fit using thermal activation model (a), Mott hopping model (b), and Efros-Shklovskii hopping model (c). For clarity, the curves have been shifted along the vertical axis, which doesn't change the fitting result (the slope of curve).

Figure 4. Schematic phase diagram describing the crossover of conduction mechanism in the $Sr_2IrO_4$ epitaxial films.

Figure 5. (a) ln$\rho$ versus $T$ of the film with thickness $t$=150 nm, which is fitted using various conduction mechanisms for different $T$ regions. (b) $w$ as a function of $T$ of the film with $t$=18 nm. Linear interpolation to the data was made before computing the derivative.

Table I. Fitting parameters to the thermal activation model and Mott variable range hopping model for epitaxial SIO films with different thicknesses.

| $t$ (nm) | $\Delta$ (meV) | $T_{M1}$ ($10^6$K) | $T_{M2}$ ($10^6$K) | $a_1$ (Å) | $a_2$ (Å) | $R_{M1}/a$ ($K^{-1/4}$) | $R_{M2}/a$ ($K^{-1/4}$) | $E_{M1}$ (meV) | $E_{M2}$ (meV) |
|---|---|---|---|---|---|---|---|---|---|
| 9 | | 60 | 131.8 | 0.6 | 0.5 | $33.0/T^{1/4}$ | $40.2/T^{1/4}$ | $1.89\ T^{3/4}$ | $2.31\ T^{3/4}$ |
| 18 | | 3.8 | 25.8 | 1.5 | 0.8 | $16.6/T^{1/4}$ | $26.7/T^{1/4}$ | $0.95\ T^{3/4}$ | $1.53\ T^{3/4}$ |
| 45 | | 1.6 | 0.67 | 2.0 | 2.7 | $13.3/T^{1/4}$ | $10.7/T^{1/4}$ | $0.77\ T^{3/4}$ | $0.62\ T^{3/4}$ |
| 82 | 97.5 | 0.87 | 0.44 | 2.5 | 3 | $11.5/T^{1/4}$ | $9.7/T^{1/4}$ | $0.66\ T^{3/4}$ | $0.55\ T^{3/4}$ |
| 150 | 86.8 | 0.65 | 0.25 | 2.7 | 3.7 | $10.6/T^{1/4}$ | $8.4/T^{1/4}$ | $0.61\ T^{3/4}$ | $0.48\ T^{3/4}$ |
| 300 | 83.8 | 0.37 | 0.21 | 3.3 | 3.9 | $9.2/T^{1/4}$ | $8.0/T^{1/4}$ | $0.53\ T^{3/4}$ | $0.46\ T^{3/4}$ |

Table II. Fitting parameters to the Efros-Shklovskii variable range hopping model for epitaxial SIO films with different thicknesses. Data of films with $t$=18 and 45 nm can not be fitted by ES-VRH model.

| $t$ (nm) | $T_{ES}$ (K) | $R_{ES}/a$ ($K^{-1/2}$) | $E_{ES}$ (meV) | $\Delta_{CG}$ (meV) | $T_{cross}$ (K) | $T_{exp}$ (K) | $T_M/T_{ES}$ |
|---|---|---|---|---|---|---|---|
| 9 | 7675 | $21.9/T^{1/2}$ | $3.77\ T^{1/2}$ | 23.7 | 7 | 180 | 7817 |
| 82 | 803 | $7.0/T^{1/2}$ | $1.22\ T^{1/2}$ | 3.0 | 23 | 28 | 547 |
| 150 | 704 | $6.6/T^{1/2}$ | $1.14\ T^{1/2}$ | 3.2 | 32 | 44 | 355 |
| 300 | 716 | $6.7/T^{1/2}$ | $1.15\ T^{1/2}$ | 3.6 | 39 | 47 | 293 |

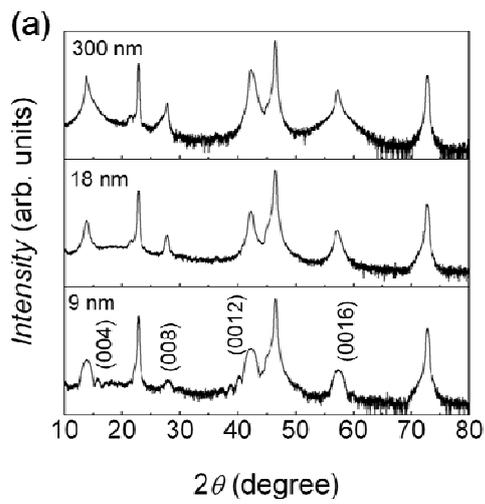
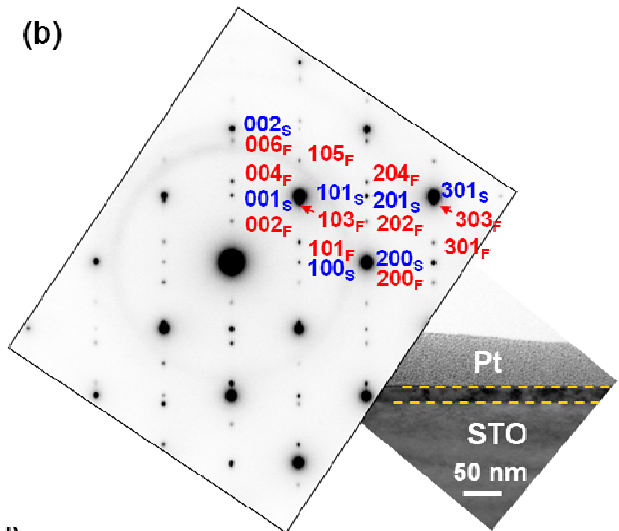
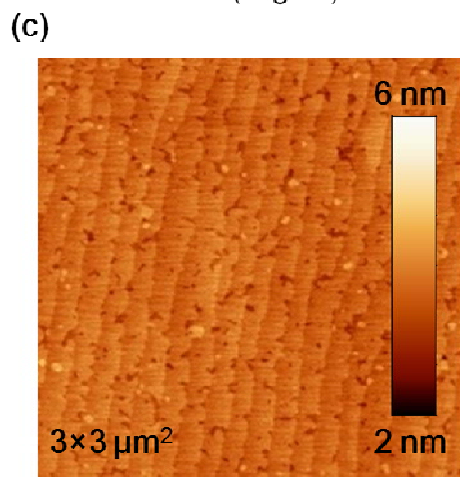
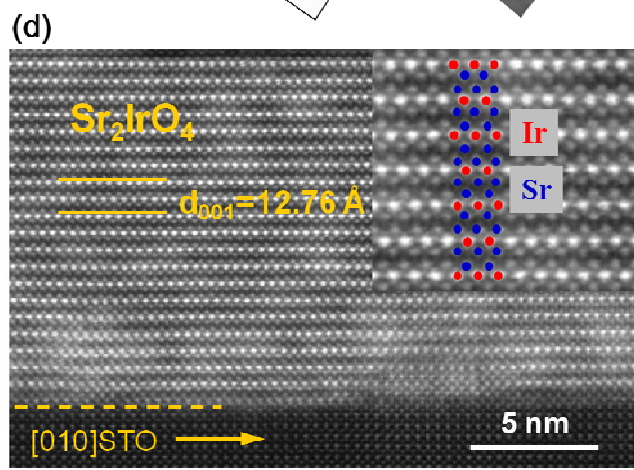

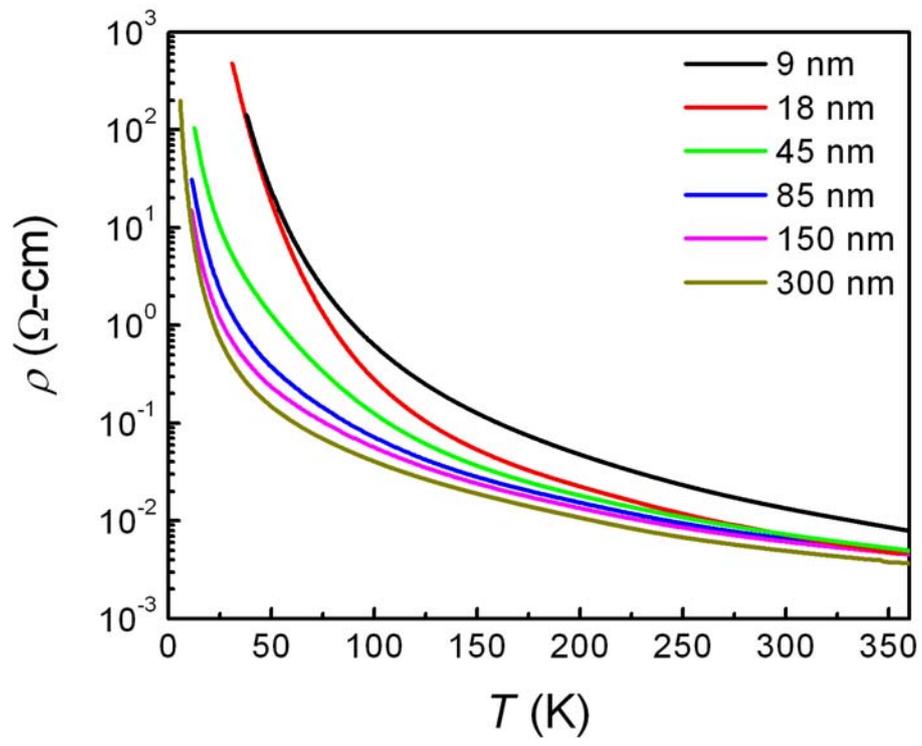

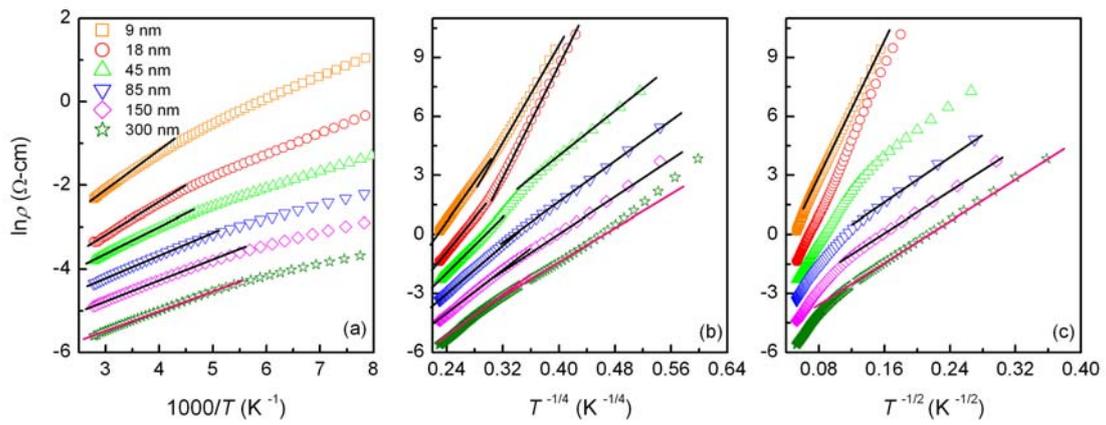

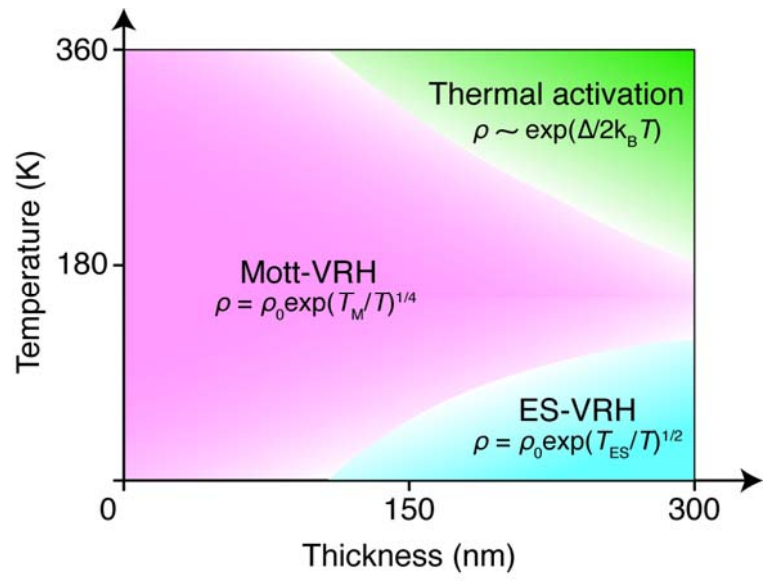

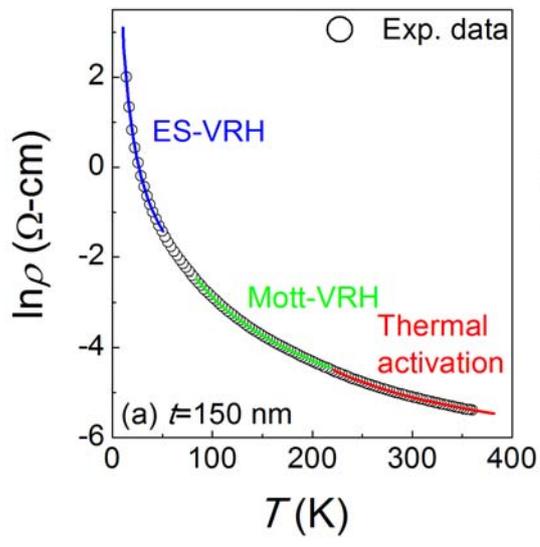 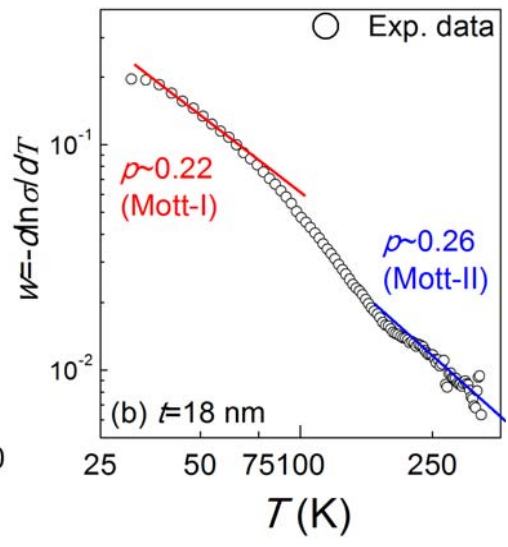